\documentclass[letterpaper]{article}
\usepackage{graphicx}
\begin{document}
\title{Scaling in the Inter-Event Time
of Random and Seasonal Systems}
\author{C\'esar A. Hidalgo R.$^{12}$\\chidalgo@nd.edu\\}
\date{}
\maketitle
\begin{center}
\textit{$^1$ Center for Complex Network Research and Department of
Physics, University of Notre Dame, Notre Dame, In. 46556\\
$^2$ Helen Kellogg Institute, Notre Dame, In, 46556}\\
\medskip

\textbf{Abstract}\\
\medskip
Interevent times have been studied across various disciplines in
search for correlations. In this paper we show analytical and
numerical evidence that at the population level a power-law can be
obtained by assuming poissonian agents with different characteristic
times, and at the individual level by assuming poissonian agents
that change the rates at which they perform an event in a random or
deterministic fashion. The range in which we expect to see this
behavior and the possible deviations from it are studied
by considering the shape of the rate distribution.
\end{center}

Power-law scaling is often considered a sign of complexity. The
independence of scale exhibited in this type of systems have
fascinated many scientists that have attempted to explain the
dynamics and correlations giving rise to this statistical property
in different systems, such as complex networks \cite{1,2,3},
fractals \cite{4} and economic fluctuations \cite{5} (for a complete
review of power-laws in nature and possible mechanisms that produce
them we encourage the reader to see \cite{newman}). This type of
function also appears in allometric laws of ecology \cite{6,7} and
in the distribution of interevent times of several different
systems. In this last context, power-law scaling has been found in
the stock exchange \cite{8,9}, earthquakes \cite{10,11} email login
times \cite{12}, print job submissions \cite{13}, email replies
\cite{14}, regular mail \cite{joao} and browsing patterns \cite{15}.
In all of these systems the distribution of interevent times scales
as $\tau^{-\alpha}$ though the exponents tend to vary from system to
system. Some of these exponents tend to be close to $\alpha=1$,
another class is close to $\alpha=3/2$ while a last class tends to
be around $\alpha=2$. The first class belongs to systems governed by
human decisions. Here it has been proposed, as a very likely
candidate, a model based on priority queues, which captures this
precise exponent \cite{14,alexei}. Whereas the second class of
behavior has been observed in the response times of Einstein's and
Darwin's corrependence\cite{joao}. Finally, the third class of
behavior has been observed in earthquakes and the stock exchange
\cite{8,9,10,11}.

In the past exponentials have been used to explain
power-laws\cite{newman}. One of the mechanism used is to consider a
variable that has an exponential distribution
\begin{equation}
f(y)\sim e^{ay}
\end{equation}
and look for a variable that is related to the first one through an
exponential, such as
\begin{equation}
x\sim e^{by}.
\end{equation}
Then, the distribution of $x$ is given by
\begin{equation}
f(x)=f(y)\frac{dy}{dx}=\frac{1}{bx}e^{(a/b) \log{(x})}=\frac{x^{a/b
-1}}{b}.
\end{equation}
This argument was first introduced by Miller \cite{miller,Li} to
explain the power law distribution of words in texts. Here we study
another simple way to extract a power-law, in this case from a
fluctuating or time dependent Poisson process. In the latter part of
this work we show that seasonal behavior can also conduct to
power-laws in the distribution of interevent times. Seasonality is a
phenomenon that becomes manifest in a variety of systems and the
change of behavior induced by it can be sufficient to abandon the
poisson paradigm.

Here we are concerned with a particular example, the distribution of
interevent, or waiting times. We argue that the $\alpha=2$ exponent
should be expected when we consider the distribution of interevent
times in a population made of several regular components which are
individually different, or when we consider individual components of
heterogeneous behavior. For the first case, we consider that the
rate that an agent performs an event in a certain time interval is
given by $p$ and the population of agents is such that we can define
$f(p)$ as the distribution of agents with particular $p's$.

For the sake of clarity we assume to have a population of agents
which send e-mails at a certain rate $p$\. We also assume that the
populations is big enough to define a distribution of rates given by
$f(p)$. We now perform a measurement in which we ask each agent for
the time elapsed between its last two e-mails and make the histogram
of this poll. The simplest case is the one in which everybody is the
same and $f(p)=\delta (p-p_0)$, where $\delta$ is Dirac's delta
distribution. In this case, the global behavior matches the
individual one. Thus the interevent time decays exponentially with a
mean given by $1/p$.

If we consider agents that have a stable individual behavior, but as
a population, have a broad distribution of rates, we would be in a
situation in which the individual behavior does not match the global
one. Individually, the agents will send emails in a Poisson fashion
allowing us to approximate their interevent times by their personal
average, which is well defined and representative at the individual
level. At the population level, we need to find the distribution of
interevent times. We can do this by simply calculating the fraction
of agents that took less than $\tau$ time units between two
consecutive emails
\begin{equation}
\label{1} P(T <\tau)=P(1/p < \tau) = P(p > 1/\tau
)=\int_{1/\tau}^{\infty}f(x)dx,
\end{equation}
and then differentiate this expression to get the probability
density
\begin{equation}
\label{result} P(T < \tau)=F(\infty)-F(1/\tau)\rightarrow
P(T=\tau)=-\frac{dF(1/\tau)}{d\tau}=f(1/\tau)\frac{1}{\tau^2}
\end{equation}
which scales as $1/\tau^{2}$ and has an envelope given by the
original function evaluated at $1/\tau$.

\begin{figure}[t]
\begin{center}
\includegraphics[width=0.9\textwidth]{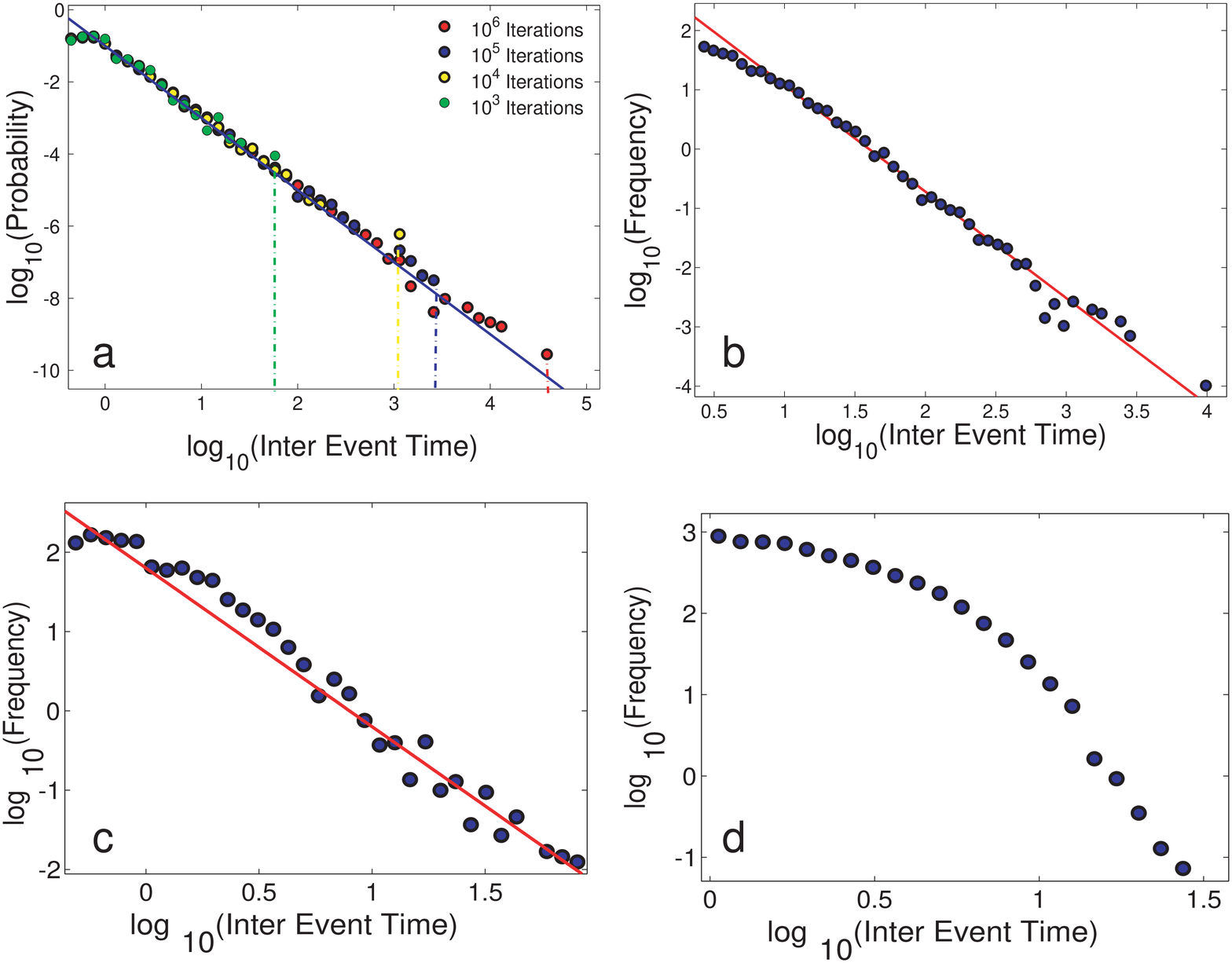}
\caption{\label{fig1} a. Finite size scaling for the distribution of
interevent times obtained when the individual probabilities of
agents was taken from a uniform distribution. The straight line has
slope -2. The dash dotted lines shows the maximum interevent time
registered for a particular number of iterations. These and all
subsequent plots were made with log-binning. b. The same result is
obtained when we consider an exponential distribution of rates given
by $f(p)=(1/8)\exp{(-8p)}$. c. When a normal distribution is chosen
for $f(p)$, the power-law behavior is present when it is wide
(N$\sim$ [ =0.4, =0.2]) and d. disappears when it is narrow (N$\sim$
[ =0.4, =0.01]).}
\end{center}
\end{figure}

Numerically, we can simulate this situation by considering a
distribution $f(p)$ and a sufficiently large population. We can do
this by picking up a particular agent with a certain $p$ from the
distribution $f(p)$ and simulate the process until it send an email
by asking at each time step if he is sending one or not. Fig.
(\ref{fig1}) shows our prediction for 3 different rate
distributions. In the case of a uniform distribution (Fig.
\ref{fig1} a), we have a system that behaves clearly as a power law
and has no envelope. We have also performed simulations with an
increasing number of agents to show that finite size scaling defines
a clear region in which this behavior is present. From equation
(\ref{result}) we have that when
\begin{equation}
f(p)\sim U[0,L]\rightarrow P(T=\tau)=1/\tau^2.
\end{equation}
and for an exponential distribution of rates we have
\begin{equation}
\label{exp1} P(T=\tau)=\frac{e^{-a/\tau}}{\tau^2}.
\end{equation}
In this case when $\tau\rightarrow\infty$, $e^{-a/\tau}\rightarrow
1$ and the behavior is the same as for the uniform distribution
which can be correctly recovered when $a\rightarrow 0$.

Deviations in the exponent can be found when the distribution of
probabilities satisfies a power law $f(p)\sim p^\beta$. Using
equation (\ref{result}) we can find that in this case
\begin{equation}
\label{power} P(T=\tau)=\tau^{-(\beta + 2)}
\end{equation}
which represents a deviation of the $\alpha=2$ exponent which occurs
in the case we inject a power-law to the system.

The studied cases do not introduce any cut-offs for large $\tau$.
This comes from the fact that long interevent times come from small
rates. All of the distributions presented above have support close
to zero, so in principle times can be infinitely long. Cut-offs in
the distribution can be introduced by restricting the support close
to zero. A simple example of this is considering the case in which
the support of $f(p)$ is restricted to the $[p_<,p_>]$. According to
the formalism presented in equations (\ref{1}) and (\ref{result}),
this introduces a hard cutoff at $\tau_{max}=1/p_<$ for large $\tau$
and at $\tau_{min}=1/p_>$ for small $\tau$. Thus we can say that as
a rule of thumb
\begin{equation}
\tau_{max} \sim 1/p_< \quad \textrm{where} \quad p_< =
min[supp[f(p)]]
\end{equation}
\begin{equation}
\tau_{min} \sim 1/p_> \quad \textrm{where} \quad p_> =
max[supp[f(p)]]
\end{equation}
We can refine this argument for $\tau_{max}$ by considering that
$f(p)$ decays in a smooth way as we approach the left edge of its
support. Approximating $f(p)$ by a power series.
\begin{equation}
f(p)=\sum_{k=0}^{\infty}A_k p^{k},
\end{equation}
and using this in equation (\ref{result}) we can conclude that for
large $\tau$ the distribution of interevent times follows
\begin{equation}
\label{condition11}
P(T=\tau)=\sum_{k=0}^{\infty}\frac{(k+1)A_k}{\tau^{k+2}}.
\end{equation}
When $\tau\rightarrow \infty$ \ref{condition11} scales as
$\tau^{k'+2}$ where $k'$ is the coefficient of the lowest order
non-vanishing expansion coefficient. To be more precise the $k'$
exponent dominates the distribution when $\tau$ satisfies
$\tau>((k+1)A_k/A_{k'})^{1/k}$ for every $k$. To simplify this
discussion, we can say that when $f(p)$ decays linearly towards zero
$\alpha=3$, and when $f(p)$ decays as a parabola we have $\alpha=4$.
The $\alpha=2$ exponent for large times is an indication that $f(p)$
can be approximated by a constant close to the left edge of its
support.

The analytical results presented above were tested via numerical
simulations. Figure (\ref{fig1} b) confirms that for an exponential
distribution of probabilities the scaling behavior is still clearly
visible and extends through several decades. We also considered the
case of a normal distribution. In this case the scaling behavior
appears when the distribution is wide enough (Fig. 1c) and
disappears for narrow bells (Fig. 1d) which have negligible support
close to zero.

So far, we have shown that we can expect a power-law for the
distribution of interevent times whenever we ask heterogeneous
individuals for the time between its last two events and poll that
data together. We have argued that the exponent should be $\alpha=2$
when the distribution of rates is broad enough and it can be
extended for several decades when $f(p)$ has support close to zero,
we have also shown that deviations from this exponent can be
explained by consideration rate distributions that scale as a power
of the rate. This argument can be extended even further to include a
population in which we do not only ask agents to tell us the
interevent time between their last two events, but we have the
distribution of events for each one of them. We are concerned with
the neutral case in which agents are not correlated in time or
across the population, and therefore they individually follow
exponential distributions. If we poll this data together by adding
up all these distributions, we also expect a power law decay with a
$\alpha=2$ exponent. This can be seen clearly by adding up
normalized exponentials representing the interevent time
distribution of regular individual agents

\begin{figure}[t]
\begin{center}
\includegraphics[width=0.9\textwidth]{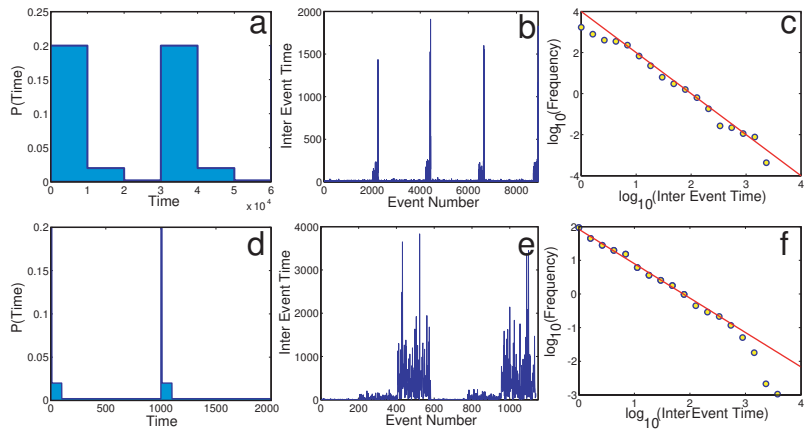}
\caption{\label{fig2} a. Periodic behavior of the rates used to
model seasonal events b. Interevent times obtained for a process
modeled using the rates in a. c. The distribution of interevent time
mimics a power-law with an exponent close to $-2$ (straight line has
slope $-2$). d. The same as a. except that in this case smaller
rates are active for longer times. e. Interevent times. f. For this
case the distribution of inter event times mimics a power-law with
an exponent close to $-1$.}
\end{center}
\end{figure}

\begin{equation}
P(T=\tau)=\sum_{i}^{n}p_i \exp{(-p_i \tau)},
\end{equation}
and assuming a uniform distribution of rates and a large enough
population we get\footnote{This method was introduced in \cite{12}
as a possible explanation for the interevent times. Although only
the case with a uniform distribution of probabilities was
considered.},
\begin{equation}
\label{integral} P(T=\tau)=\int_{0}^{\infty}p \exp{(- p \tau)}dp =
-\frac{d}{d\tau}\int_{0}^{\infty} \exp{(-\tau p)}dp,
\end{equation}
which can be easily solved resulting in
\begin{equation}
\label{13} \frac{d}{d\tau}\frac{\exp{(-\tau
p)}}{\tau}\Big|_{0}^{\infty}=-\frac{d}{d\tau}\big(\frac{1}{\tau}\big)=\frac{1}{\tau^2}.
\end{equation}

This argument can be easily generalized to include any distribution
$f(p)$. In this case we have
\begin{equation}
\label{res2} P(T=\tau)= \int_{0}^{\infty} f(p)\: p\: e^{-p\: \tau}dp
\end{equation}
which may sound redundant given equations (\ref{1}) and
(\ref{result}). In fact, we introduce both methodos because the
first one is easier to use in some cases in which the integral given
in equation (\ref{res2}) is not trivial to calculate. The conceptual
difference of the two methods for calculating interevent time
consists that in the first one we assume that the times coincide
precisely with their expected values in the exponential
distributions ($1/p$) whereas in the second one we consider all
possible values with associated with a given probability. In the
case that $f(p)$ is an exponential distribution normalized in the
$[0,1]$ interval we have that
\begin{equation}
P(T=\tau)= \frac{1}{(a+\tau)^2}.
\end{equation}
whereas when $f(p)$ is a power-law this second method requires us to
solve
\begin{equation}
P(T=\tau)=\int_{0}^{1}p^{\beta+1}e^{-p\tau}dp
\end{equation}
which can be expressed in terms of a $\Gamma$ function as
\begin{equation}
P(T=\tau)=\bigg( \frac{1}{\tau}\bigg)^{\beta+2}\Gamma(\beta+2)
\end{equation}
being hardly more useful than equation (\ref{power}).

The arguments used so far have been used to show that at a
population level we are likely to observe a power-law distribution
of interevent times when we poll up a large population of
non-identical users. So far we have assumed that single users
perform events at a fixed rate and act accordingly. But what can we
expect if we allow individual agents to vary? Going back to our
email user analogy, we can imagine that we measure the time
intervals between several consecutive emails for a particular agents
and discover that the histogram of interevent times is a power law
with an exponent $\alpha=2$. We can show that in this case, an
uncorrelated random process can also explain the scaling exponent.
For this matters, let us consider an agent that initially send
emails at a rate $p_0$. After sending an email, we record the
interevent time and start waiting again. If $p_0$ is fixed, the
interevent time will decay exponentially, but if we allow this rate
to change in time this would not be necessarily true. The simplest
case would be to let $p_0$ evolve in a purely random fashion, in
other words, after the agent sends an email, we randomly draw a new
rate $p_1$ from the $[p_<,p_>]$ interval. If this is the case, this
system would be the same that the one in which we consider several
agents with different rates, and therefore, it is again obvious to
expect an $\alpha=2$ exponent in the interevent time distribution.
Thus, an agent that varies its behavior in a random way maps the
same model that a population of significantly different users. The
way in which an individual agent varies its behavior is actually not
important, as long as different rates are chosen. In fact, we can
relax the assumption that rates vary randomly and instead choose a
periodic function for them. In figure (\ref{fig2}) we show an
example of this process in which we simulate a system in which the
event execution rate changes from 0.2 to 0.02 to 0.002 to then be
reseted back to its original value of 0.2 to start all over again.
Here the rate depends only on time and does not change after an
event is registered (figure \ref{fig2} a). The interevent times are
therefore correlated and the system changes periodically from
activity to inactivity (figure \ref{fig2} b). The distribution of
interevent times mimics a power-law with $\alpha=2$, but upon closer
look, one can identify three humps which coincide with the expected
times of the three fixed probabilities involved in the process. In
order to mimic a power-law with seasonality one needs to consider
$p(t)$ such that the values taken by this function are widely
distributed\footnote{we mean widely in a logarithmic sense. A
similar number of values per order of magnitude.} and that the
function has regions in which it varies slowly enough (or not at
all) to allow interevent times to be consistent with the rates
proper corresponding to each time.\footnote{An exaggeration of this
is to consider a two step function with values 0.1 and 0.01 that has
a period of two time steps. In this case it is obvious that the
system is going to be dominated by $p=0.1$ because the period of the
function is shorter than the expected time associated with
$p=0.01$.} Figures (\ref{fig2}) a,b and c; correspond to a case in
which $p(t)$ mimics an exponential distribution, this is because
after a linear time it decays an order of magnitude. We can shift
the exponent in this case by working with a $p(t)$ that mimics a
power law. As an example we show the case in which we consider the
same three probabilities as before (0.2, 0.02 and 0.002) but instead
of lasting the same amount of time each, we make them last $100$,
$1000$ and $10000$ time steps respectively. In this case the longer
interevent times are as frequent as the shorter ones and the system
mimics a power law with $\alpha=1$ (Figure (\ref{fig2}) f).

In the light of the previous results and examples we have shown
cases in which we can obtain power-laws by combining exponentials.
The cutoff of the power-law depends on the minimum of the support of
the distribution $f(p)$ and the exponent for large $\tau$ depends on
the leading term of the power series expansion of $f(p)$. In the
case that the function approaches the left edge of its support as a
constant we expect the exponent for large $\tau$ to be $\alpha=2$,
but in the case that the function approaches this edge as a function
with singular support, we have a decrease on the exponent which is
equal to the order of the singularity. Finally when the exponent
approaches the edge as a polynomial $\alpha$ increases by the degree
of the of the $k'$ term of the polynomial for
$\tau>((k+1)A_k/A_{k'})^{1/k} \forall k$.

In the case of large earthquakes, it has been shown that the
interevent time of earthquakes larger than a given magnitude scales
as a power-law with a $\alpha=2$ exponent \cite{10}. Omori's law is
not valid for the interevent time between earthquakes larger than a
certain magnitude, which is the case in which you see this exponent.
This is because Omori's law deals with the aftershocks which are
clearly correlated. It was also argued that the $\alpha=2$ exponent
indicates a time correlated behavior, because the interevent time
distribution is not Poissonian. From the analytical arguments shown
above, we can see that the $\alpha=2$ exponent is precisely what is
expected for the uncorrelated case in which we consider that an
earthquake occurs with a probability that is randomly reset after
each event. In the case of the stock exchange \cite{9}, it was shown
that the scaling exponent tends to decrease as the threshold on the
normalized fluctuations increases. In other words, when waiting
times between large fluctuations are considered, the scaling
exponent approaches $\alpha=1$ indicating possibly a different
mechanism like a queue model\cite{Alexeilast} or a power-law
injection, whereas when small variations are considered the exponent
is close to $\alpha=2$, which could be a signature of seasonality
\cite{8} or uncorrelated probability variations.

Despite the simplicity of our calculations, Poisson process are
usually assumed in stochastic modeling. Here we have shown that when
we have a broad population which participates in individual poisson
process or agents which have non-stationary behavior the waiting
time distributions follows a power-law. In the case of a simple
poisson processes the first thing that should be considered is that
usually not all of the components of the system behave in the same
way, this consideration alone is enough to define a scaling region
that has not usually been considered.

I would like to thank D. Menzies, A.-L. Barab\'asi and the Referees
for helpful reviews of this manuscript. I would also like to
acknowledge funding from NSF ITR 0426737, NSF ACT/SGER 0441089 and
the James S. McDonnell Foundation.

\end{document}